\documentclass[10pt]{article} 
\usepackage[preprint]{tmlr}
\usepackage{graphicx} 
\usepackage{multirow} 
\usepackage{booktabs}
\usepackage[dvipsnames]{xcolor}
\usepackage{pifont}

\usepackage{amsmath,amsfonts,bm}

\def\eqref#1{equation~\ref{#1}}

\def\1{\bm{1}}

\DeclareMathAlphabet{\mathsfit}{\encodingdefault}{\sfdefault}{m}{sl}
\SetMathAlphabet{\mathsfit}{bold}{\encodingdefault}{\sfdefault}{bx}{n}

\usepackage{hyperref}
\usepackage{url}

\usepackage{subcaption}   
\usepackage{caption}

\title{Leveraging Reference Documents for Zero-Shot Ranking \\ via Large Language Models}

\author{
  Jieran Li\textsuperscript{1},
  Xiuyuan Hu\textsuperscript{1},
  Yang Zhao\textsuperscript{1},
  Shengyao Zhuang\textsuperscript{2},
  Dongbiao Sun\textsuperscript{1}
  Hao Zhang\textsuperscript{1} \\
  \textsuperscript{1}Department of Electronic Engineering, Tsinghua University \\
  \textsuperscript{2}CSIRO\\
  lijr23@mails.tsinghua.edu.cn\\
}

\def\month{MM}  
\def\year{YYYY} 
\def\openreview{\url{https://openreview.net/forum?id=XXXX}} 

\begin{document}

\maketitle

\begin{abstract}

Large language models (LLMs) have proven strong zero-shot rerankers, yet the two dominant paradigms expose a sharp accuracy-efficiency trade-off. Existing methods mainly fall into two categories: Individual-scoring (pointwise) issues $O(n)$ parallel calls but suffers from calibration drift across isolated prompts; Comparative-sorting (pairwise/listwise) alleviates drift via explicit inter-document comparison, but incurs higher-than-linear inference or long single-call latency. To address their limitations, we propose \textbf{RefRank}, a reference-anchored framework that marries the throughput of Individual-scoring with the calibration benefits of Comparative-sorting. RefRank prompts the LLM to score each candidate relative to a fixed anchor document harvested from the first-stage top-k list; all candidates are thus implicitly compared through the same anchors while parallelism is preserved. The method is training-free, adds no extra model calls, and keeps complexity at O(n). Across six standard benchmarks and multiple backbones, RefRank significantly outperforms Individual-scoring baselines and surpasses Comparative-sorting competitors with only negligible overhead.

\end{abstract}

\section{Introduction}

Modern large-scale search pipelines universally adopt a retrieve-then-rerank architecture~\citep{karpukhin2020dense}. The first-stage retriever is required to be both fast and memory-efficient, so that millions of passages can be scanned in milliseconds~\citep{robertson2009probabilistic}.
These systems typically return the top-$k$ ($k\!=\!100$--$1\,000$) candidate passages, which are subsequently fed to a second-stage reranker that performs heavy cross-attention or generation to produce a more accurate ordering before the final answer is synthesized~\citep{nogueira2019multi,karpukhin2020dense,lassance2024splade}.
Large language models~(LLMs) such as GPT-3~\citep{brown2020language}, PaLM~\citep{chowdhery2023palm}, Llama~\citep{touvron2023llama}, and Flan-T5~\citep{wei2021finetuned} have shown strong zero-shot reranking ability simply by prompting them with a query–passage pair and asking for a relevance decision~\citep{yates2021pretrained, agrawal2023large, kojima2022large, wang2023can}.
This immediately turns LLMs into a viable reranking option that requires no training data, and they are now being embedded in production RAG stacks where latency, throughput, and calibration directly impact end-to-end user experience.~\citep{li2024retrieval}.

Two ranking paradigms have emerged, exposing a fundamental accuracy-efficiency trade-off. \textbf{Individual-scoring}~(pointwise) asks the LLM to emit an independent relevance score for every candidate; the scores are then sorted to produce the final ranking~\citep{liang2022holistic, zhuang2023beyond, zhuang2023open, sachan2022improving, guo2025mcranker}.
Because each passage is processed in isolation, the method is parallel and easy to batch, yielding high throughput on both GPU and CPU hardware.
Unfortunately, recent work repeatedly shows that LLM scores suffer from severe calibration drift: semantically similar passages receive different priors, degrading fidelity when subtle relevance distinctions matter~\citep{zhuang2023beyond, wang2023can}.
To mitigate calibration error, researchers have turned to \textbf{Comparative-sorting}~(pairwise, setwise, listwise) prompts that explicitly ask the LLM to compare two or more passages and output their relative order~\citep{qin2023large, zhuang2024setwise, pradeep2023rankvicuna, sun2023chatgpt}.
While inter-document comparison significantly improves discriminative power, it incurs higher query complexity:
ranking $n$ passages requires $O(n^2)$ pairwise calls~\citep{chen2025tourrank}, whereas listwise approaches need only one call but consume $O(n)$ input tokens, leading to prolonged inference latency, higher memory footprints, and degraded attention fidelity as $n$ grows.
At web scale, comparative rerankers therefore become latency-bound, memory-bound, and harder to parallelize,  largely offsetting the throughput benefits that modern accelerators otherwise provide~\citep{rivera2022continuous, zhuang2023beyond}.

In this paper we ask: \textit{Can we retain the parallel efficiency of pointwise scoring while still injecting the calibration benefits of inter-document comparison?}  
We answer the question affirmatively with \textbf{RefRank}, a reference-anchored relative scoring framework. Instead of requesting an individual score, the LLM is prompted to emit a relative relevance score with respect to a single anchor passage.
The anchor is selected from the top-k list returned by any first-stage retriever—be it BM25, dense retriever, or hybrid—turning the initial ranking into free supervision ~\citep{xu2017query}.
Because every passage is compared to the same reference, the resulting scores are globally calibrated; yet each document still requires only one forward pass, preserving full parallelism.

RefRank's only wrinkle is anchor sensitivity: picking the anchor from different ranks shifts the score distribution enough to perturb the final order.
Following the pseudo-relevance intuition that the first retrieved hit is already query-relevant~\citep{xu2017query}, we treat the top-1 passage as a free, high-quality anchor and implement two variants.
\textbf{RefRank-Single} scores every candidate once against that single anchor, while \textbf{RefRank-Multiple} averages the relative scores obtained against the top-4 first-stage hits; neither variant adds training or extra prompt text.
On TREC-DL 2019 with Flan-T5-XL and 100 candidates the single-anchor version finishes in 1.9\,s per query---60$\times$ faster than pairwise Bubblesort (115.8\,s), 58$\times$ faster than listwise (111.7\,s) and 7.7$\times$ faster than Setwise-Heapsort (14.7\,s)---issuing exactly 100 forward calls, the same cost as pointwise baselines.
Across six test collections (TREC-DL 2019, 2020 and four BEIR subsets), Flan-T5-XL/XXL with RefRank-Multiple yields the highest average NDCG@10 (0.614/0.616), and even Llama-3.1-8B reaches 0.598, matching the best reported zero-shot figure and confirming that the gain is backbone-agnostic.

The contributions of this paper can be summarized as follows:
\begin{enumerate}
    \item \textbf{RefRank}, a reference-anchored relative-scoring framework that retains the parallelism of pointwise rerankers while achieving comparative-level accuracy.
    \item A training-free method to convert first-stage retrieval rankings into anchor supervision, instantiated as two variants: RefRank-Single and RefRank-Multiple.
    \item RefRank establishes a new state-of-the-art NDCG@10 on TREC-DL 2019 and yields statistically significant improvements across six datasets, while adding negligible latency.
\end{enumerate}


\section{Related Works}

\begin{figure}[ht]
  \centering
  \begin{subfigure}[b]{\linewidth}
    \includegraphics[width=\linewidth]{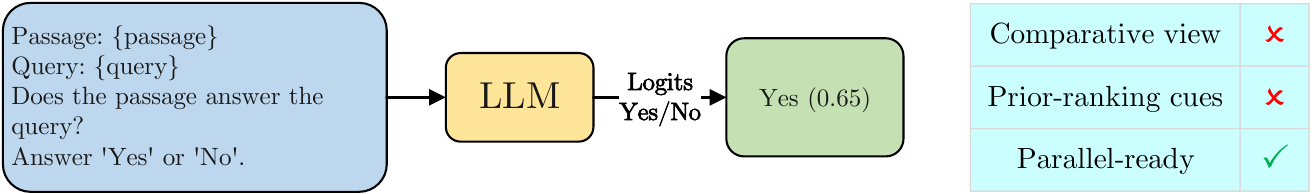}
    \caption{Individual scoring }
  \end{subfigure}\\[1.5mm]

  \begin{subfigure}[b]{\linewidth}
    \includegraphics[width=\linewidth]{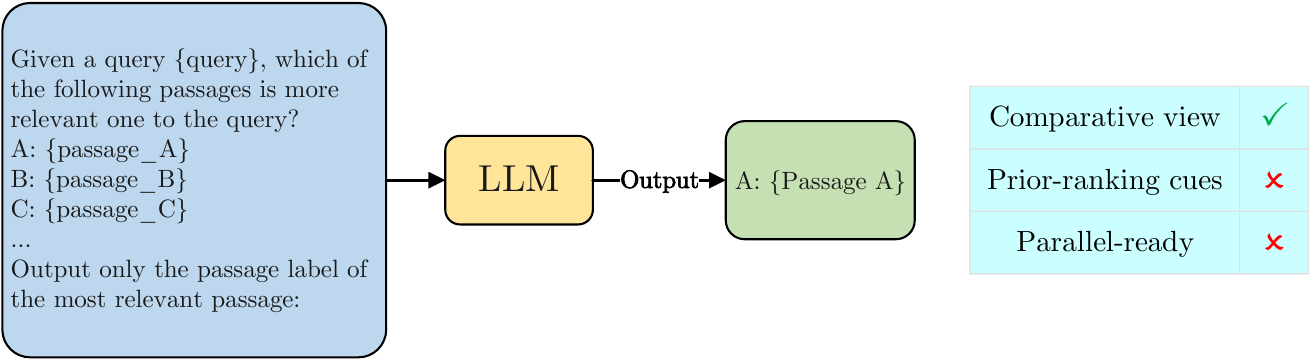}
    \caption{Comparative sorting}
  \end{subfigure}\\[1.5mm]

  \begin{subfigure}[b]{\linewidth}
    \includegraphics[width=\linewidth]{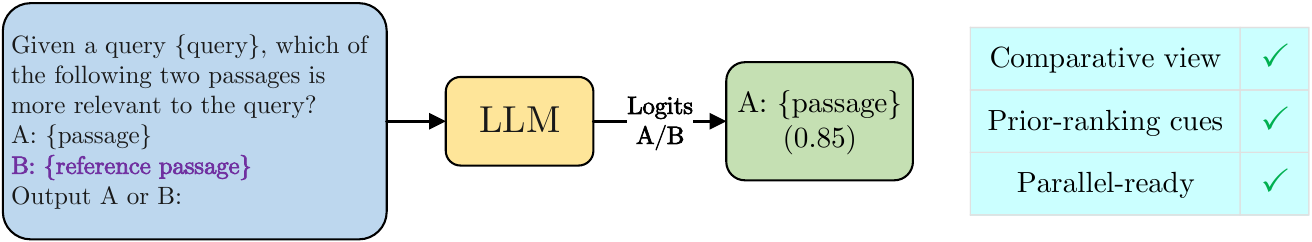}
    \caption{\textbf{Our RefRank}}
  \end{subfigure}

  \caption{Paradigms of zero-shot reranking: (a) Individual-scoring, (b) Comparative-sorting, and (c) \textbf{Our RefRank}.}
  \label{fig:zero_shot_compare}
\end{figure}

Existing LLM-based zero-shot rerankers follow two mutually exclusive paradigms:
(i)~\textbf{Individual-scoring} (pointwise) and
(ii)~\textbf{Comparative-sorting} (pairwise/setwise/listwise).
Figure~\ref{fig:zero_shot_compare} situates \textbf{RefRank} with respect to these lines of work.

\subsection{Individual-Scoring}
Pointwise methods quantify the relevance of a single document $d$ to query $q$ in isolation.
They fall into two sub-families:
\textbf{(a) Direct relevance assessment} feeds the $(q,d)$ pair into the LLM and interprets the log-probability of the token ``yes'' as the relevance score \citep{liang2022holistic,zhuang2023beyond}.
The prompt can be extended to k-class labels (e.g., Not Relevant, Somewhat Relevant, Highly Relevant) to obtain finer-grained scores \citep{zhuang2023beyond}.
\textbf{(b) Query-generation similarity} first asks the LLM to produce candidate queries from $d$, then measures the semantic similarity between the generated queries and the original $q$ \citep{zhuang2023open}.
This requires two separate LLM calls per document.
Because documents are scored independently, pointwise approaches can be batched and scale linearly.
Their fatal weakness is calibration drift: the model lacks an internal zero-point, so scores across documents are incomparable and fine-grained ranking suffers.

\subsection{Comparative-Sorting}

\textbf{Pairwise} prompts compare two documents $(d_i,d_j)$ and ask the LLM to return the more relevant one ~\citep{qin2023large}.
Enumerating all ordered pairs requires $O(n^2)$ inference calls; even with adaptive or sampling-based sorting, the number of sequential LLM invocations remains super-linear and quickly becomes the latency bottleneck for web-scale candidate pools ~\citep{gienapp2022sparse,mikhailiuk2021active,chen2025tourrank}.
\textbf{Setwise} extends pairwise to small subsets of size $c$.
Heap-sort or tournament procedures lower the call count, yet each prompt is longer and the process is still inherently sequential ~\citep{zhuang2024setwise,yoon2024listt5,podolak2025reproducibilityadvancingzeroshotllm}.
\textbf{Listwise} feeds the entire candidate list into the LLM and requests a ranked output~\citep{pradeep2023rankvicuna,sun2023chatgpt} or the log-probability of special rank tokens~\citep{reddy2024first}.
Context length grows linearly with~n, so latency and memory scale accordingly; sliding-window tricks~\citep{sun2023chatgpt} alleviate but do not remove this growth, and position bias further limits usable list sizes.

In short, the field is locked in a trilemma:
pointwise sacrifices discrimination for speed;
pairwise/setwise sacrifice parallelism for accuracy;
listwise sacrifices memory for a single pass.
Critically, none of the existing paradigms recycle the high-quality ranking signals that an upstream retriever has already produced.
\textbf{RefRank} breaks this impasse by converting the top-$k$ hits of any first-stage run into a shared, in-pool anchor.
Relative scoring against this single anchor yields well-separated, inter-calibrated logits in one parallel forward pass---no quadratic calls, no context explosion, no sequential bottleneck.
Thus, RefRank is the first LLM reranker to deliver the speed of pointwise, the discriminative power of pairwise, and the single-call convenience of listwise, while simultaneously exploiting upstream retrieval cues rather than ignoring them.

\section{Method}

\subsection{Problem Statement and Design Goals}
\label{sec:problem}
Let \( q \) denote a natural-language query and \( \mathcal{D} = \{d_1, d_2, \dots, d_n\} \) represent a candidate pool of \( n \) passages retrieved by an arbitrary first-stage retriever \( \mathcal{R}_0 \), such as BM25, dense retrieval, or hybrid methods.  
The initial ranking is denoted by \( \pi_0 \), where \( \pi_0(i) \) returns the index of the passage ranked at position \( i \).  
The objective is to generate a new permutation \( \pi^* \) that minimizes the expected ranking risk under the latent relevance distribution, subject to the following constraints:

\begin{enumerate}
  \item \textbf{Zero-shot}: No task-specific training or gradient updates are permitted.
  \item \textbf{Linear complexity}: The inference cost must scale as \( O(n) \) with respect to the number of candidates.
  \item \textbf{Fully batchable}: LLM inferences must be independent and executable in a single GPU/CPU batch.
  \item \textbf{High effectiveness}: The method should achieve competitive ranking accuracy (e.g., NDCG@10) without leveraging additional supervision signals.
\end{enumerate}

\textbf{Reference-Anchored Relative Scoring.}  
Existing paradigms either score each passage independently or perform direct pairwise comparisons across the entire candidate set.  
In contrast, the proposed approach reuses the high-quality signal already present in the initial ranking \( \pi_0 \) by selecting a small subset of top-ranked passages as shared anchors.  
By requesting LLM to express a relative preference against a fixed anchor, additive model priors are effectively canceled, yielding globally calibrated scores with only one forward pass per candidate.

\subsection{RefRank-Single: A Linear-Time Contrastive Scorer}
\label{sec:single}

\begin{figure}[ht]
  \centering
  \begin{subfigure}[b]{0.46\linewidth}
    \includegraphics[width=\linewidth]{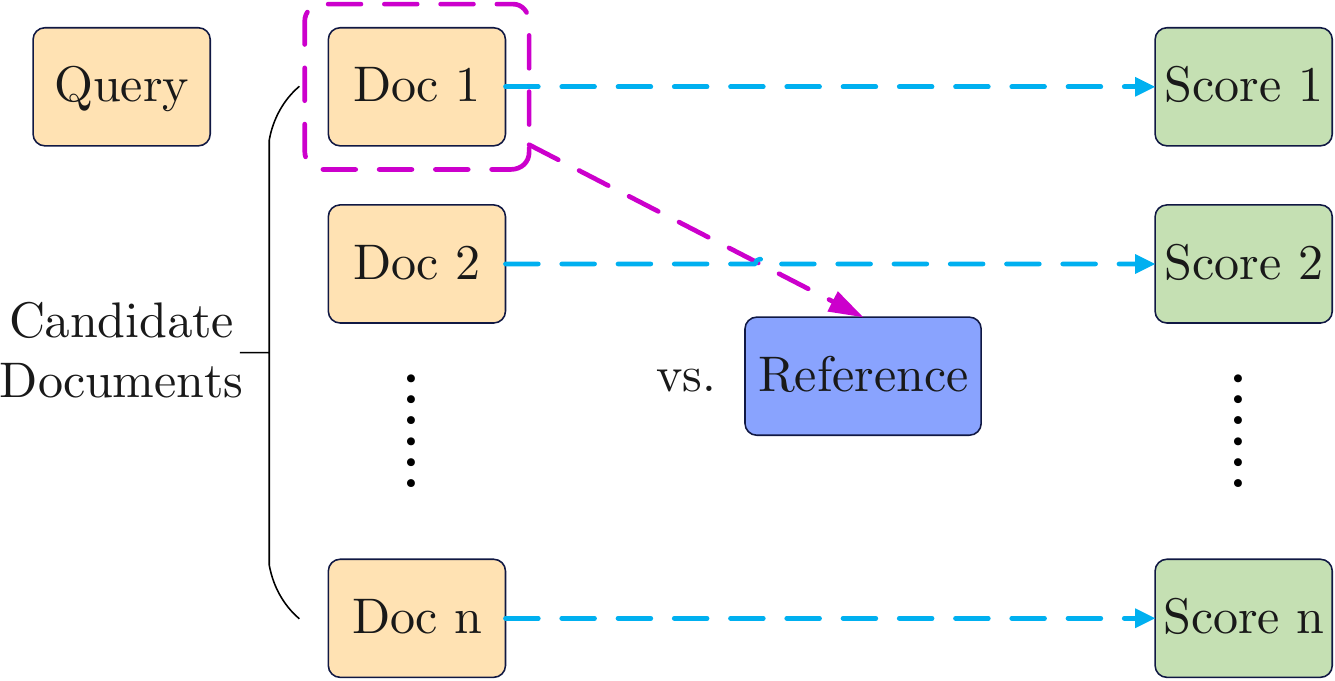}
    \caption{RefRank-Single}
  \end{subfigure}
  \hfill
  \begin{subfigure}[b]{0.52\linewidth}
    \includegraphics[width=\linewidth]{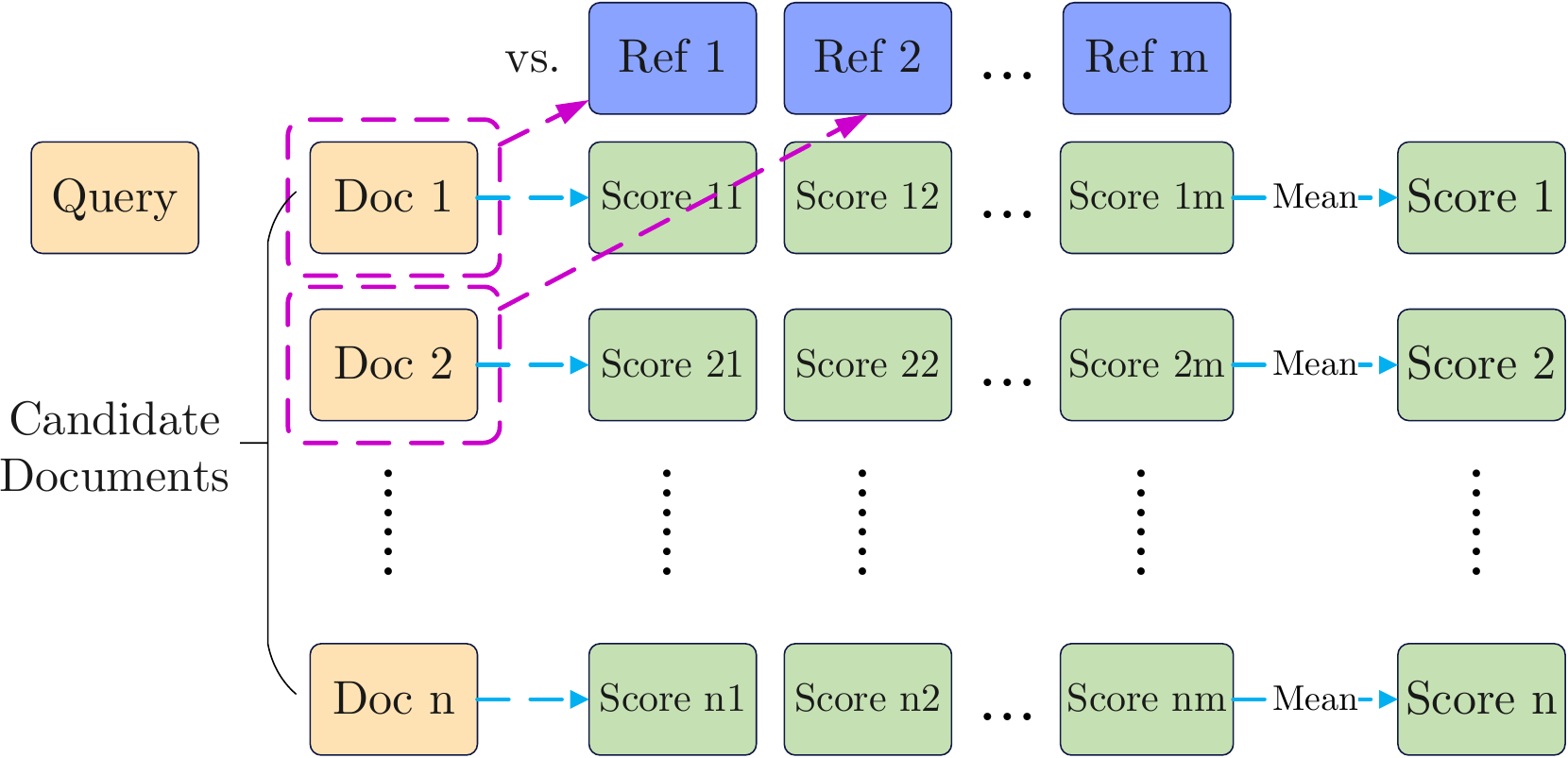}
    \caption{RefRank-Multiple}
  \end{subfigure}

  \caption{Illustration of RefRank's two inference schemes. 
  {\bf (a)}~RefRank-Single: each candidate is compared against one fixed reference (e.g.,top-1) and ranked by the resulting relative score.  
  {\bf (b)}~RefRank-Multiple: each candidate is scored against \emph{top-$k$} references and the final score is obtained by mean pooling, yielding smoother and more robust rankings.}
  \label{fig:refrank}
\end{figure}

An overview of the proposed variants is illustrated in Figure~\ref{fig:refrank}.  
\textbf{RefRank-Single} contrasts each candidate against a single shared anchor (e.g., the top-1 passage), whereas \textbf{RefRank-Multiple} aggregates log-odds against the top-$k$ anchors.

Formally, let the reference passage be \( d_r = d_{\pi_0(1)} \), i.e., the highest-ranked passage under \( \pi_0 \).  
For each candidate \( d_i \in \mathcal{D} \), the frozen LLM \( \mathcal{M}_\theta \) is prompted to perform a single-token comparative classification:
\begin{align}
    p_i &= P_{\mathcal{M}}\!\left(\text{``}d_i \succ d_r\text{''} \mid q, d_i, d_r\right), \\[2pt]
    p_r &= P_{\mathcal{M}}\!\left(\text{``}d_r \succ d_i\text{''} \mid q, d_i, d_r\right).
\end{align}

The relevance score of \( d_i \) is defined as the log-odds ratio:
\begin{equation}
    s(d_i; q, d_r) = \log p_i - \log p_r,
    \label{eq:score}
\end{equation}
which neutralizes constant model priors and yields a zero-centered real-valued score.  
The final ranking is obtained by sorting the set \( \{s(d_i; q, d_r)\}_{i=1}^n \) in descending order.  
The procedure invokes exactly \( n \) forward passes, each consuming two passages and the query, thereby ensuring linear complexity.

\textbf{Prompt Template.}  
To ensure reproducibility, a deterministic prompt is employed, as depicted in Figure~\ref{fig:zero_shot_compare}.  
The query and the two passages are concatenated with explicit delimiters.  
The LLM is instructed to output a single token from \( \{\texttt{A}, \texttt{B}\} \), where \texttt{A} corresponds to \( d_i \) and \texttt{B} corresponds to \( d_r \).  
Token probabilities are extracted from the softmax layer and mapped to \( p_i \) and \( p_r \) accordingly.

\begin{figure}[ht]
  \centering
  \begin{subfigure}[b]{0.49\linewidth}
    \includegraphics[width=\linewidth]{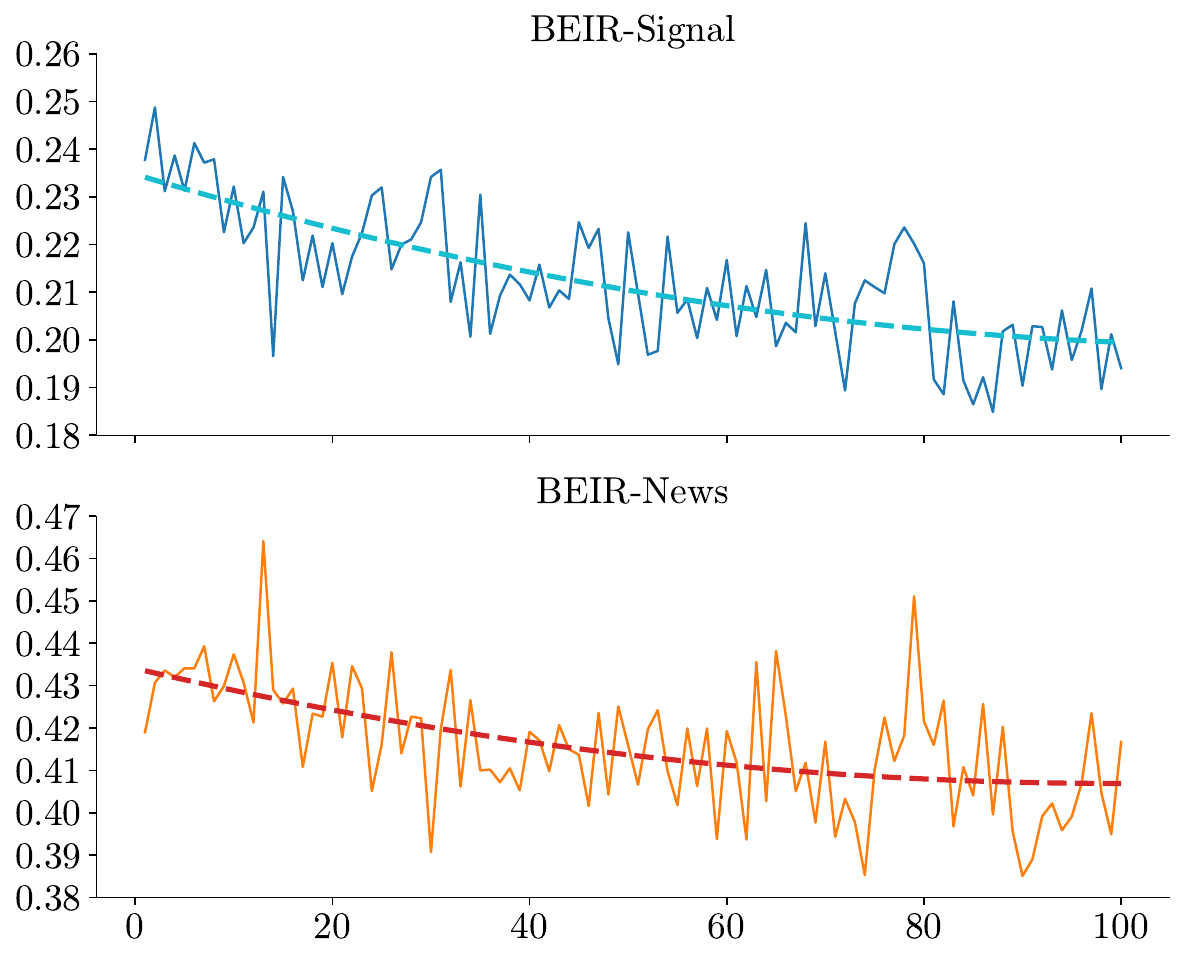}
    \caption{Llama-3.1-8B}
  \end{subfigure}
  \hfill
  \begin{subfigure}[b]{0.49\linewidth}
    \includegraphics[width=\linewidth]{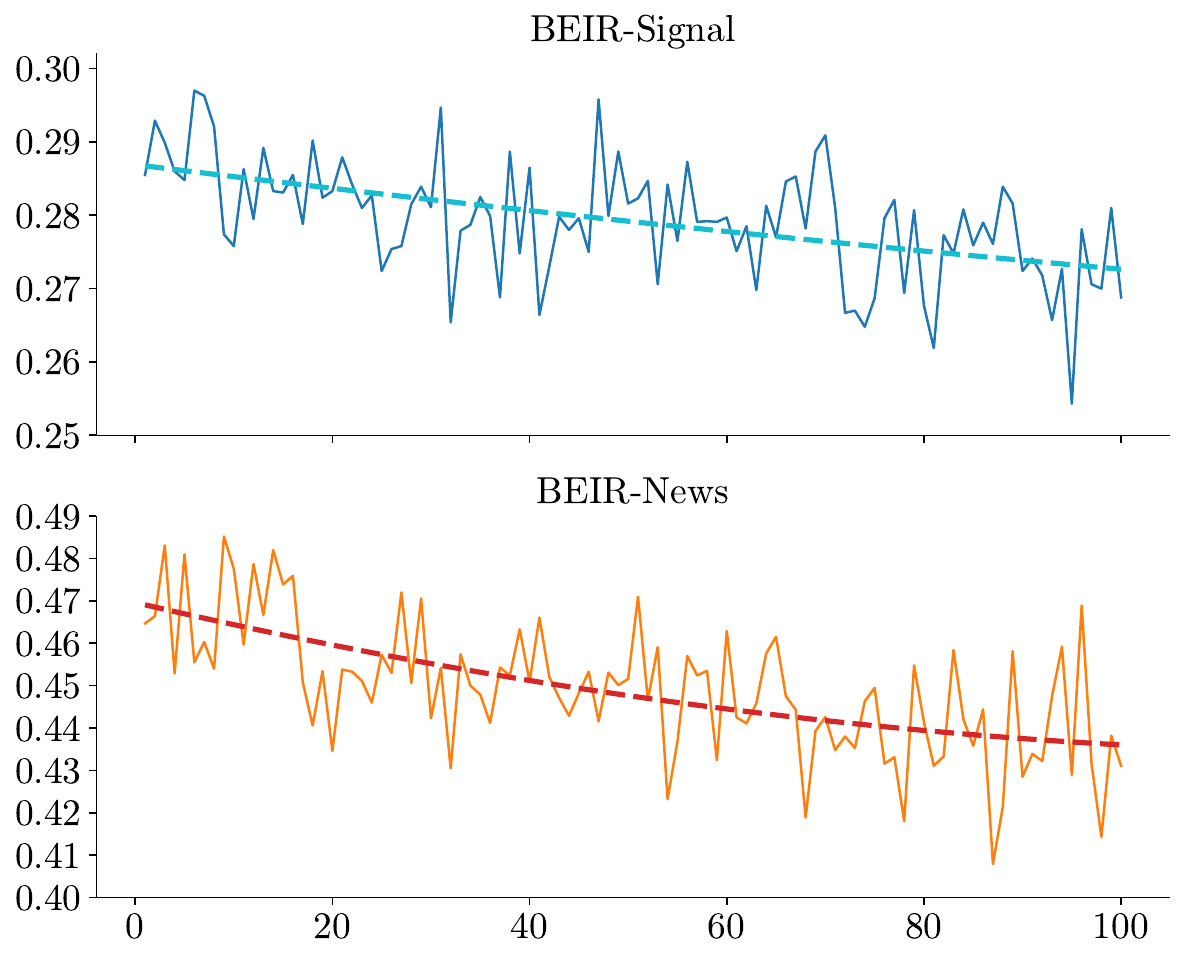}
    \caption{Flan-T5-XL}
  \end{subfigure}

  \caption{NDCG@10 versus selected reference passage index for two models.}
  \label{fig:scores}
\end{figure}

\textbf{Empirical Validation of Anchor Position.}  
To investigate the impact of anchor-passage selection on retrieval effectiveness, we conduct an empirical analysis on two BEIR subsets---Signal and News---using two off-the-shelf LLMs: Flan-T5-XL (3B) and Llama-3.1-8B.  
For each query we sequentially treat the $k$-th passage ($k=1,\dots,100$) as the reference passage $d_{r}$ and measure the corresponding NDCG@10.  
As shown by the dashed line in Figure~\ref{fig:scores}, performance degrades as $k$ increases, indicating that lower-ranked passages provide weaker supervision.

\begin{figure}[ht]
  \centering
  \begin{subfigure}[b]{0.49\linewidth}
    \includegraphics[width=\linewidth]{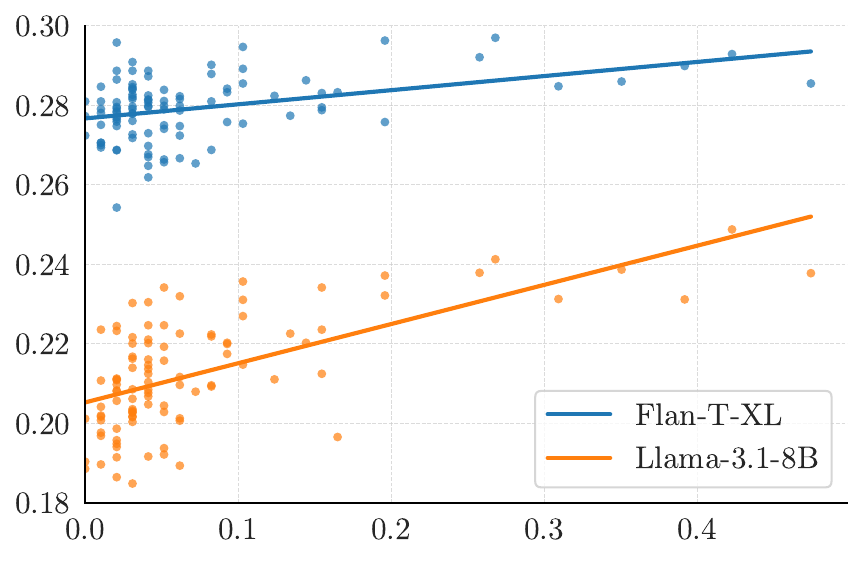}
    \caption{BEIR-Signal}
  \end{subfigure}
  \hfill
  \begin{subfigure}[b]{0.49\linewidth}
    \includegraphics[width=\linewidth]{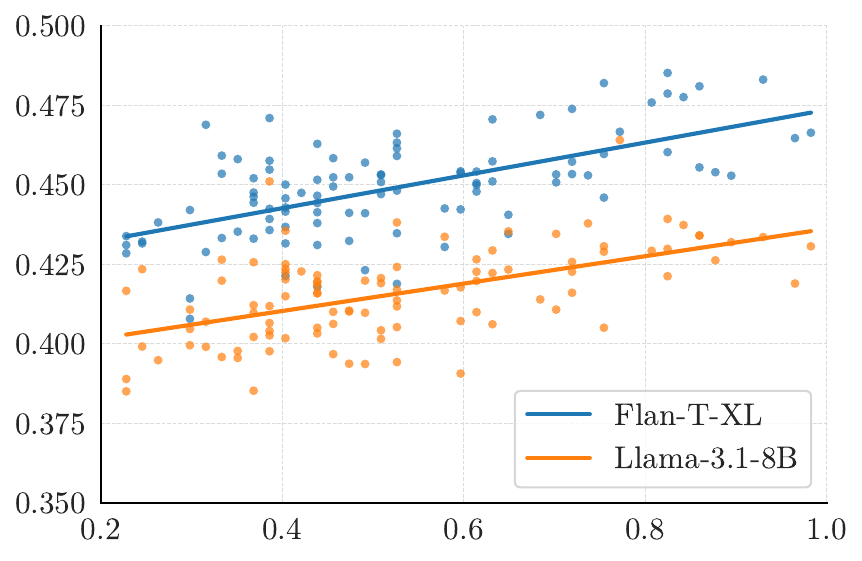}
    \caption{BEIR-News}
  \end{subfigure}

  \caption{Correlation between the fraction of reliable reference passages and NDCG@10. The fitted regression line indicates a strong positive relationship.}
  \label{fig:fraction}
\end{figure}

\textbf{Reliability-score analysis.}
To further understand this phenomenon, we introduce an additional metric: the reliability fraction of the selected passage.  
Formally, let $m$ be the number of test queries and $T_{k}$ the number of queries whose selected passage is reliable (i.e., contains the correct answer).  
The fraction of reliable reference passages is defined as
$\text{Reliability@}k = \frac{T_{k}}{m}$.
We plot Reliability@$k$ against the observed NDCG@10 in Figure~\ref{fig:fraction}.  
A strong positive correlation emerges: higher reliability consistently yields higher NDCG@10, corroborating that selecting a reliable anchor passage is crucial for effective supervision.
Notably, passages ranked higher usually exhibit greater RReliability@$k$.

Collectively, our dual analysis—via NDCG@10 trends and passage reliability—demonstrates that
(i)~the top-1 passage consistently delivers the highest-quality supervision signal, and
(ii)~passages within the top-5 remain competitively reliable.
Consequently, we adopt $d_{\pi_{0}(1)}$ as the \emph{default} reference passage, while retaining early-position anchors (top-5) as an optional hyper-parameter for practitioners.

\subsection{RefRank-Multiple: Robust Aggregation via Top-k Anchors}
\label{sec:multiple}

A single reference may introduce noise or topical bias.  
To mitigate this, ~\eqref{eq:agg} is generalized to an ensemble of anchors sampled from the high-confidence region of \( \pi_0 \).  
Let \( R = \{d_{\pi_0(1)}, \dots, d_{\pi_0(k)}\} \) denote the set of top-$k$ passages under \( \pi_0 \).  
For each \( d_i \in \mathcal{D} \), the score \( s(d_i; q, d_r) \) is computed for every \( d_r \in R \), and the final score is obtained by averaging:
\begin{equation}
    S(d_i; q, R) = \frac{1}{k} \sum_{d_r \in R} s(d_i; q, d_r).
    \label{eq:agg}
\end{equation}

The final ranking is produced by sorting \( \{S(d_i; q, R)\}_{i=1}^n \).  
The procedure requires \( k \cdot n \) forward passes, which remains \( O(n) \) for any constant \( k \).

\begin{figure}[ht]
  \centering
  \begin{subfigure}[b]{0.49\linewidth}
    \includegraphics[width=\linewidth]{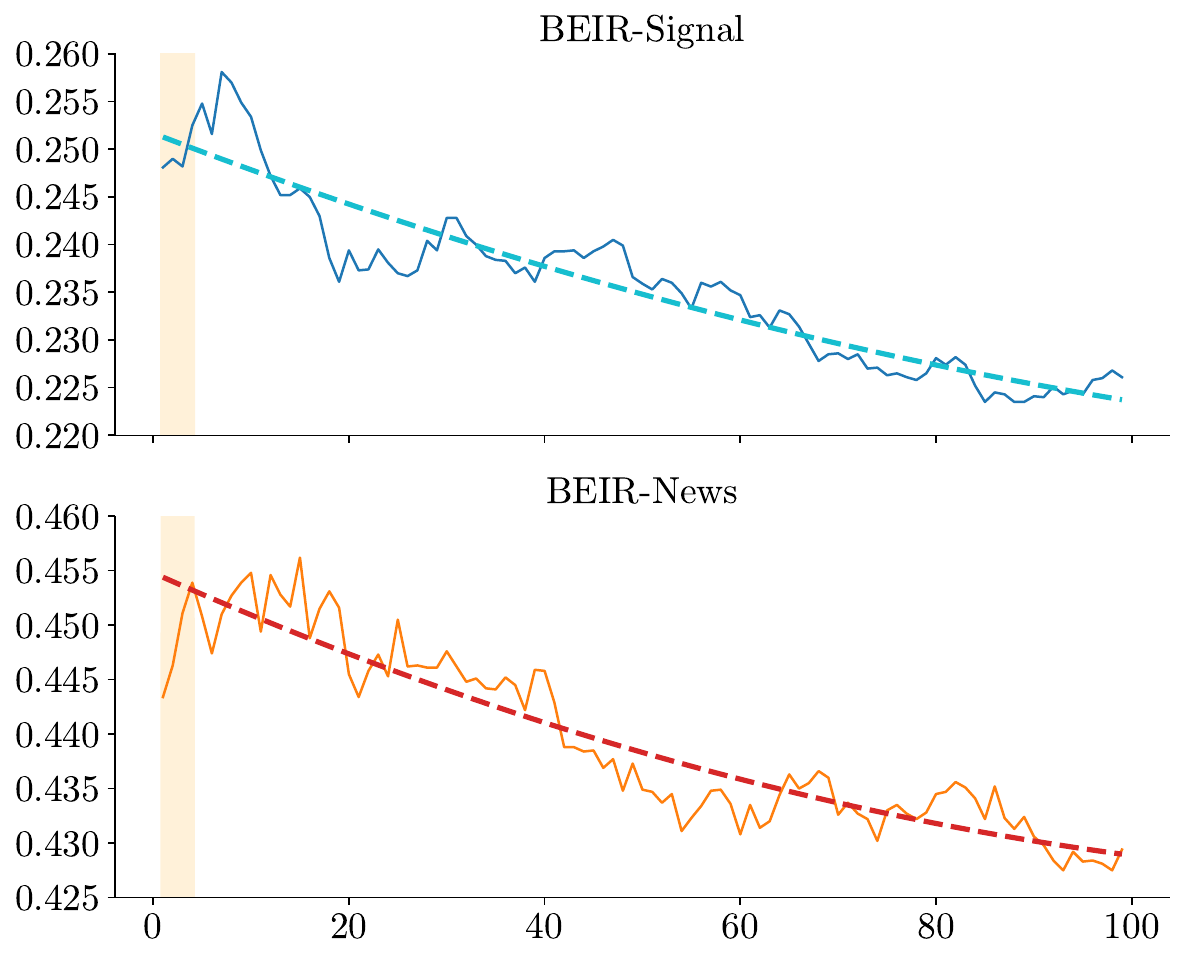}
    \caption{Llama-3.1-8B}
  \end{subfigure}
  \hfill
  \begin{subfigure}[b]{0.49\linewidth}
    \includegraphics[width=\linewidth]{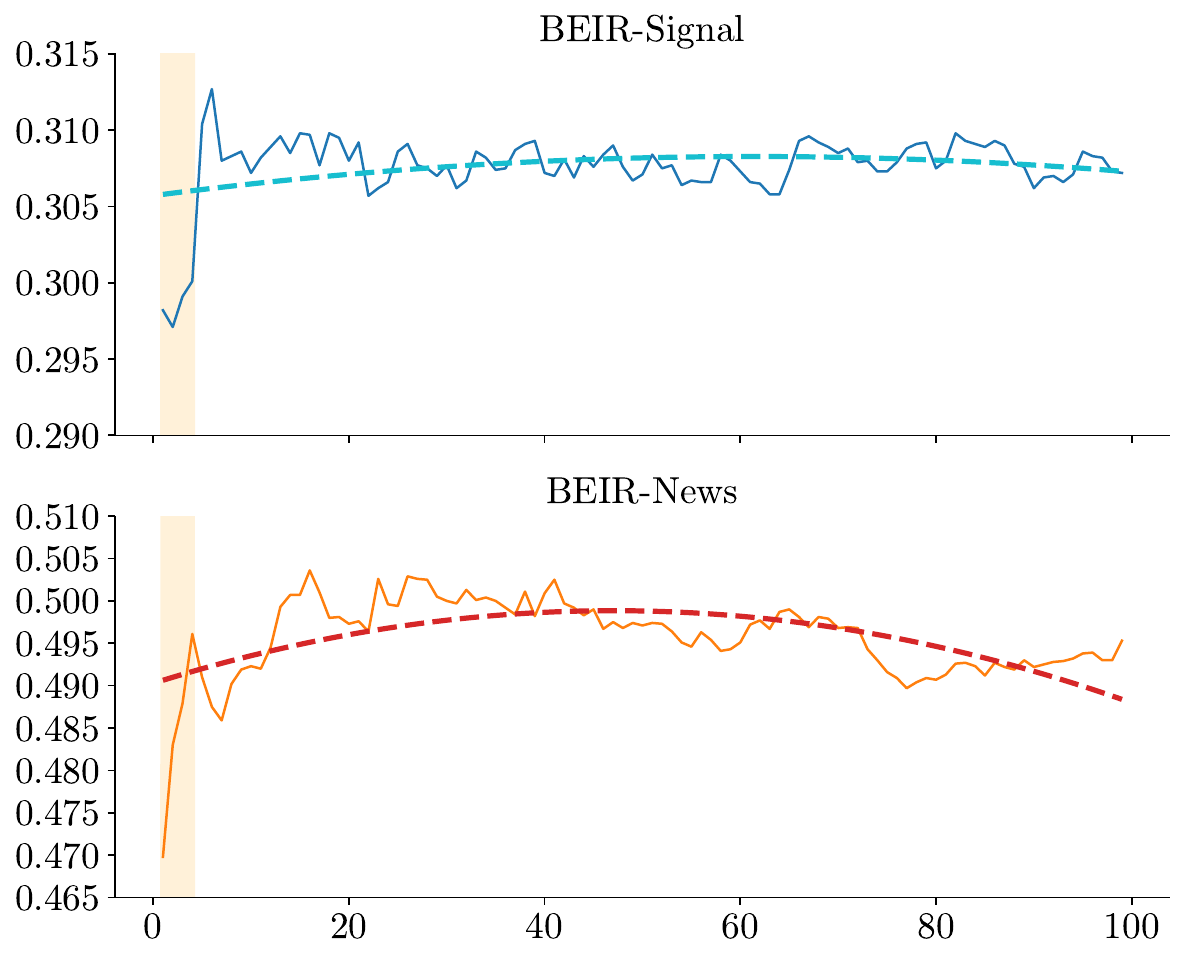}
    \caption{Flan-T5-XL}
  \end{subfigure}

  \caption{NDCG@10 versus the number of top-$k$ reference documents sequentially weighted.}
  \label{fig:weights}
\end{figure}

\textbf{Choice of \( k \) and Anchor-Sampling Strategy.}  
The pilot experiment is repeated while varying the number of anchors \( k \) from 2 to 100.  
Figure~\ref{fig:weights} exhibits an increase-then-decrease pattern: NDCG@10 improves until \( k \approx 10 \), after which it either plateaus or mildly deteriorates.  
This confirms that a compact set of high-quality anchors is sufficient.  
To further quantify the importance of anchor quality, two strategies are compared for assembling \( R \) when \( k = 4 \):
\begin{itemize}
  \item \textbf{Sequential:} The top-4 passages under \( \pi_0 \), i.e., \( d_{\pi_0(1)}, \dots, d_{\pi_0(4)} \).
  \item \textbf{Random:} Four passages sampled without replacement from the top-100 list.
\end{itemize}

\begin{figure}[ht]
  \centering
  \begin{subfigure}[b]{0.49\linewidth}
    \includegraphics[width=\linewidth]{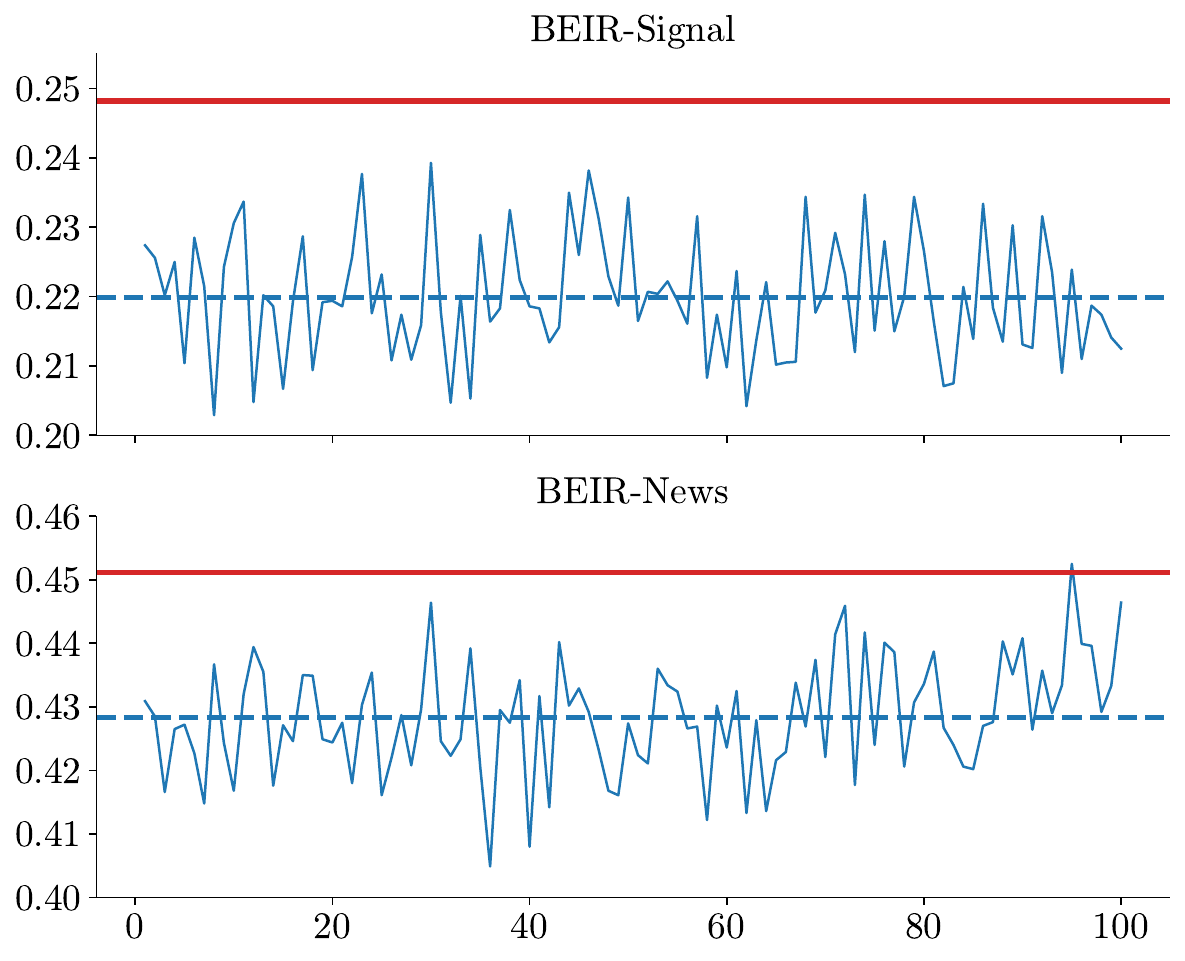}
    \caption{Llama-3.1-8B}
  \end{subfigure}
  \hfill
  \begin{subfigure}[b]{0.49\linewidth}
    \includegraphics[width=\linewidth]{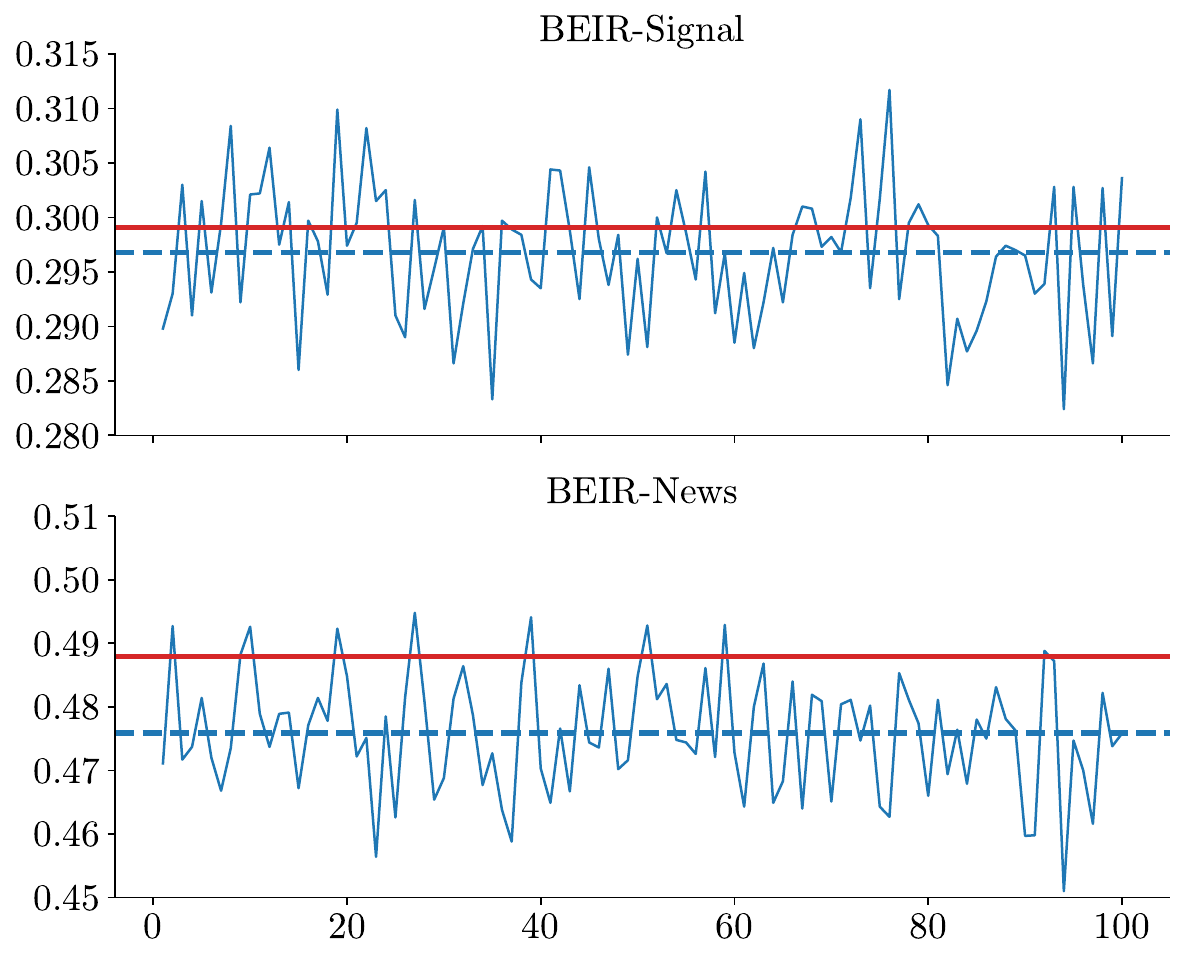}
    \caption{Flan-T5-XL}
  \end{subfigure}

  \caption{NDCG@10 of RefRank using the top-4 first-stage documents as references (solid line) versus randomly sampled 4-document references (dashed line).}
  \label{fig:random}
\end{figure}

Figure~\ref{fig:random} summarises the results obtained on both the Signal and News datasets.
For both Llama-3.1-8B and Flan-T5-XL, the mean NDCG@10 achieved by the random strategy (blue dashed line) is consistently lower than that of the sequential strategy (red solid line), demonstrating that the latter’s superiority is systematic rather than an artefact of a few lucky queries.
Indeed, the red solid curve lies entirely above the blue dashed curve, and—most notably—for Llama-3.1-8B the sequential strategy outperforms nearly every single one of the 100 independent random runs (blue solid line).

\subsection{Complexity and Efficiency Analysis}
\label{sec:complexity}

We evaluate five complementary desiderata for zero-shot LLM rerankers: \textbf{Comparative signal}: Whether the method leverages pairwise or setwise comparisons. \textbf{Initial ranking}: Whether the method exploits the initial ranking \( \pi_0 \). \textbf{GPU batching}: Whether all inference calls are independent and amenable to batching. \textbf{Number of calls}: The total number of LLM forward passes in the worst case. \textbf{Input length}: The number of passages per prompt, where \( L \) denotes the length of a single passage.

Table~\ref{tab:tab1} summarizes the asymptotic behavior of representative methods, where n denotes the number of candidates, k is the early-stopping threshold for top-k retrieval, r gives the number of listwise repetitions, s is the sliding-window size, c is the branching factor in the setwise tournament, and L is the number of tokens in a single passage.

\begin{table}[ht]
  \centering
  \caption{Asymptotic comparison of zero-shot LLM rerankers.
  \textbf{Comparative}: uses pairwise/setwise comparison.
  \textbf{Sorting}: exploits initial $\pi_{0}$.
  \textbf{Batching}: independent calls amenable to batching.
  \textbf{Number}: total forward calls (worst-case).
  \textbf{Length}: input passages per call ($L$ denotes one passage).)}
  \label{tab:tab1}
  \begin{tabular}{l|c|c|c|c|c}
    \toprule
    Method & Comparative & Sorting & Batching & Number & Length \\
    \midrule
    Pointwise-RG & \textcolor{red}{\ding{55}} & \textcolor{red}{\ding{55}} & \textcolor{Green}{\ding{51}} & $O(n)$ & $L$ \\
    Pairwise-Bubblesort & \textcolor{Green}{\ding{51}} & \textcolor{red}{\ding{55}} & \textcolor{red}{\ding{55}} & $O(kn)$ & $2L$ \\
    Listwise & \textcolor{Green}{\ding{51}} & \textcolor{red}{\ding{55}} & \textcolor{red}{\ding{55}} & $O(r(n/s))$ & $sL$ \\
    Setwise-Heapsort & \textcolor{Green}{\ding{51}} & \textcolor{red}{\ding{55}} & \textcolor{red}{\ding{55}} & $O(k \log_c n)$ & $cL$ \\
    \textbf{RefRank} & \textcolor{Green}{\ding{51}} & \textcolor{Green}{\ding{51}} & \textcolor{Green}{\ding{51}} & $O(n)$ & $2L$ \\
    \bottomrule
  \end{tabular}
\end{table}

\textbf{Pointwise-RG} scores each $d_i$ independently; no comparison, no use of~$\pi_{0}$, but fully batchable.
\textbf{Pairwise-Bubblesort} performs $O(kn)$ comparisons; every call is a pair, prohibiting large batches because the next pair depends on the previous decision.
\textbf{Listwise} ranks windows of $s$~passages and slides; length grows as~$sL$, while calls scale with~$r(n/s)$.
\textbf{Setwise-Heapsort} builds a tournament tree; although logarithmic in calls, each prompt contains $c$~full passages.
\textbf{RefRank} issues only n forward passes—one per candidate—each containing the query and the same anchor passage; this keeps memory constant, latency linear in n, and lets all calls run in parallel for full accelerator utilisation.

\section{Experiments}
\label{sec4}

\subsection{Datasets and evaluations}

In order to assess the effectiveness and efficiency of our research, we utilized several standard evaluation datasets in information retrieval: the TREC-DL 2019 and 2020, along with the BEIR benchmark datasets. These datasets facilitate empirical analysis by providing a robust framework for evaluation. To maintain the consistency of assessment, all query results were generated using a BM25-based initial retrieval method \cite{lin2021pyserini}. From these results, we selected the top 100 candidate documents for subsequent reranking.
The TREC datasets serve as established benchmarks in the realm of information retrieval. Specifically, we focused on two subsets: TREC-DL 2019 and 2020. All queries are derived from the MS MARCO v1 corpus, a comprehensive resource containing approximately 8.8 million passages.
In addition to the TREC datasets, we incorporated the BEIR datasets, which encompass a diverse array of information retrieval tasks across multiple domains. In our analysis, we selected Covid, NFCorpus, DBPedia and SciFact.
The datasets are selected to enable a nuanced evaluation of retrieval effectiveness across a variety of contexts. 

We adopt NDCG@10 as the primary evaluation metric, as it is the official standard for these benchmarks. This choice facilitates direct comparison with prior work and enhances the credibility of our experimental findings.
To assess efficiency, we employ the following metrics:
\textbf{Average number of LLM inferences per query.} Reranking 100 documents often requires multiple inferences due to the input length constraints of LLMs. A higher number of inferences leads to increased computational cost, making this a key efficiency indicator.
\textbf{Average number of prompt tokens per query.} This metric reflects the total number of input tokens required to rerank 100 documents. Since the computational cost of transformer self-attention scales with input length, larger prompts incur higher overhead. Thus, prompt token count serves as an important measure of efficiency.
\textbf{Average query latency.} We measure runtime efficiency as the mean per-query latency on a single GPU, with queries processed sequentially. For methods that support batching, we use the maximum feasible batch size to fully utilize GPU capacity. While cross-query batching may be employed in practice for non-batching methods, evaluating such engineering-level optimizations is beyond the scope of this paper.

In our study, we utilized the standard configuration of the Pyserini Python library to generate preliminary BM25 ranking results for all experimental datasets \citep{lin2021pyserini}. We conducted a systematic evaluation of three models, including Flan-T5-Xl (3B), Flan-T5-XXl (11B)\citep{wei2021finetuned} and Llama-3.1-8B. All experimental method settings should be kept consistent with ~\cite{zhuang2024setwise}.
 We carried out the efficiency evaluations on a local GPU workstation equipped with an AMD EPYC 7742 64-Core CPU, a NVIDIA DGX A800 GPU with 80GB of memory.

\subsection{Baseline}

To ensure a comprehensive evaluation, we compare \textbf{RefRank} against four representative zero-shot reranking baselines:

\textbf{Pointwise-RG} Following the Relevance Generation (RG) approach \citep{liang2022holistic}, we prompt the LLM with a query--document pair and estimate a relevance score based on the model's binary (\emph{yes}/\emph{no}) response. Documents are then ranked by these scores.

\textbf{Pairwise-Bubblesort} We adopt the pairwise prompts introduced by \cite{qin2023large} and instantiate the comparison strategy with a bubble-sort algorithm. Each swap decision is delegated to the LLM, leading to an \(O(kN)\) inference complexity.

\textbf{Listwise} Consistent with the list generation framework, we utilize the prompt template proposed by \cite{sun2023chatgpt}. The LLM receives a sliding window of \(s\) documents and outputs their relative ordering; the process is repeated for \(r\) rounds to cover the full candidate list.

\textbf{Setwise-Heapsort} Building on the set-selection strategy of \cite{zhuang2024setwise}, we employ a heap-sort mechanism. At each step the LLM identifies the most relevant document from a set of \(c\) candidates, yielding an \(O(k\log_{c}N)\) inference count.

By systematically categorizing and evaluating these contrasting methods—pointwise, pairwise, listwise, and setwise—we provide a holistic view of current zero-shot reranking techniques and their respective trade-offs between effectiveness and efficiency.

\subsection{Results and Analysis}
\label{sec:results}

\subsubsection{Ranking Efficiency}
\label{sec:efficiency}

Table~\ref{tab:tab2} summarises both effectiveness (NDCG@10) and efficiency metrics on the TREC-DL 2019 test set.
We report: (1) the number of LLM forward calls per query;
(2) total prompt tokens (query + passages);
(3) average end-to-end latency with maximal GPU batching on one NVIDIA A800;
and (4) NDCG@10.
\begin{table*}[ht]
  \centering
  \caption{Effectiveness and efficiency on TREC-DL 2019.  
Efficiency metrics: (1) number of LLM inferences, (2) total prompt tokens per query, (3) average latency per query on a single NVIDIA A800 with maximal batching. \underline{\textbf{Bold-underlined}}: best; \underline{underlined}: second-best.}
  \label{tab:tab2}
  \begin{tabular}{l|l|cccc} \toprule
    LLM & methods & Inferences  & Tokens & Latency(s) & NDCG@10 \\
    \midrule
    - & BM25 & -  & - & - & 0.506 \\
    \midrule
    \multirow{5}{*}{Flan-T5-XL} & Pointwise-RG & \underline{\textbf{100}}  & \underline{\textbf{16k}} & \underline{\textbf{1.4}} & 0.650 \\
    & Pairwise-Bubblesort & 887 & 400k  & 115.8 & 0.683 \\
    & Listwise & 245  & 119k & 111.7 & 0.568 \\
    & Setwise-Heapsort & 130  & 41k & 14.7 & \underline{0.692}  \\
    & \textbf{RefRank-Single(1)} & \underline{\textbf{100}}  & \underline{27k} & \underline{1.9} & \underline{\textbf{0.694}} \\
    \midrule
    \multirow{5}{*}{Flan-T5-XXL} & Pointwise-RG & \underline{\textbf{100}}  & \underline{\textbf{16k}} & \underline{\textbf{6.1}} & 0.642 \\
    & Pairwise-Bubblesort & 870 & 394k & 329.3 & 0.679 \\
    & Listwise & 245  & 119k & 173.2 & 0.660 \\
    & Setwise-Heapsort & 130  & 42k & 45.3 & \underline{0.706}  \\
    & \textbf{RefRank-Single(1)} & \underline{\textbf{100}}  & \underline{27k} & \underline{8.3} & \underline{\textbf{0.707}} \\
    \midrule
    \multirow{5}{*}{Llama-3.1-8B} & Pointwise-RG & \underline{\textbf{100}}  & \underline{\textbf{19k}}  & \underline{\textbf{4.7}} & 0.617 \\
    & Pairwise-Bubblesort & 852 & 411k & 161.4 & 0.652 \\
    & Listwise & 245  & 142k & 156.3 & \underline{0.681} \\
    & Setwise-Heapsort & 126  & 42k & 19.1 & 0.662  \\
    & \textbf{RefRank-Single(1)} & 100  & \underline{29k} & \underline{5.7} & \underline{\textbf{0.683}} \\
    \bottomrule
  \end{tabular}
\end{table*}

\textbf{BM25} is included as a lexical anchor; its NDCG@10 of 0.506 remains well below all method runs, confirming the need for semantic reranking.
\textbf{Pointwise-RG} consumes the minimal budget: exactly $n\!=\!100$ calls, $\sim$16k tokens, and sub-second latency (1.4 s for Flan-T5-XL).
However, its NDCG@10 (0.650) is significantly lower than that of \textbf{RefRank-Single} (0.694), showing that the reference signal improves quality without practically increasing overhead.

\textbf{Pairwise-Bubblesort} requires output-dependent comparisons; as a result it issues 850--900 calls and 400k tokens, two orders of magnitude more than RefRank.
Its internal sorting loop is inherently sequential, disabling full batching and driving latency to 115--330 s.
NDCG@10 (0.679--0.683) remains below RefRank despite the 28--60$\times$ compute increase.
\textbf{Listwise} reduces the number of calls to 245, but it still produces 119k--142k tokens and incurs 111--173\,s latency.
Its NDCG@10 (0.568--0.681) is unstable across models.
\textbf{Setwise-Heapsort} issues 126--130 calls and 42k tokens; latency drops to 14--45\,s, yet remains 3--7$\times$ slower than RefRank.
More importantly, the tournament tree introduces stage-level sequential dependencies that prohibit single-batch execution.
Consequently, its NDCG@10 (0.662--0.706) is competitive but still significantly below that of RefRank on every backbone.

\textbf{RefRank-Single} keeps the same number of calls (100) and only adds one extra passage per prompt (reference), yielding $\sim$27k tokens (second-lowest).
Latency rises marginally to 1.9 s (+0.5 s), remaining within the same order of magnitude as Pointwise.
Crucially, because every prompt is independent, the entire candidate set can be scored in a single GPU batch; hence, wall-clock time grows linearly with $n$ and is bounded by memory bandwidth, not by sequential dependencies.
Across three backbones, RefRank consistently delivers the \textbf{best} NDCG@10 (0.694--0.707).

In summary, \textbf{RefRank} is the only reranking method that simultaneously delivers four desirable properties: it issues exactly n inference calls (linear cost), keeps token overhead constant at one additional passage regardless of model size, allows all calls to be batched for full-GPU utilisation, and still achieves state-of-the-art effectiveness.

\subsubsection{Ranking Effectiveness}
Table~\ref{tab:tab3} reports NDCG@10 on TREC-DL 2019/2020 and four BEIR subsets.
\textbf{RefRank-Multiple(4)} consistently delivers the highest macro-average across all three LLMs:
0.610~(Flan-T5-XL),
0.617~(Flan-T5-XXL) and
0.598~(Llama-3.1-8B),
outperforming the strongest competitor on the Flan-T5 family by +0.4--1.2\,pp absolute.
\textbf{RefRank-Single(1)} already improves upon its Pointwise-RG baseline by
+11.6\,\%--13.7\,\% relative, confirming that a single high-quality reference is universally beneficial.

\textbf{Encoder--decoder architecture (Flan-T5)}
On both XL and XXL variants, the six-dataset average exhibits a stable method ordering:
RefRank-Multiple(4) $>$ RefRank-Single(1) $\approx$ Pairwise-Bubblesort $>$ Setwise-Heapsort $>$ Listwise $>$ Pointwise-RG.
This monotonic sequence shows that converting the initial retrieval set into an explicit anchor document yields larger and more consistent gains than escalating the complexity of pairwise or setwise comparisons.
Scaling from XL to XXL enlarges the RefRank-Multiple(4) margin from +0.4\,pp to +1.2\,pp, verifying stable improvement with model size.

\textbf{Decoder-only architecture (Llama-3.1-8B)}
Thanks to its native long-context capacity, Listwise reaches the same macro-average as RefRank-Multiple(4) (0.598), making the two methods jointly best.
Nevertheless, RefRank-Multiple(4) still produces the highest single-run scores on DL-19 (0.704) and DL-20 (0.606) while operating 27$\times$ faster than Listwise.
This suggests that the decoder-only architecture can partially close the gap through extended context and auto-regressive scoring, yet RefRank retains an edge in both efficiency and head-to-head effectiveness on query sets that benefit from explicit reference comparison.

\textbf{Single~vs.~Multiple references}
Ensembling four references boosts macro-average NDCG@10 by only +0.5–2.5\,\% relative to one reference, indicating that a single well-chosen document already encodes the majority of comparative signal; the multi-reference variant acts primarily as a safeguard for ambiguous queries. In summary, RefRank raises the ceiling of zero-shot reranking without sacrificing efficiency, reconciling the effectiveness of pairwise/listwise schemes with the speed of pointwise methods and providing a practical drop-in upgrade for production systems.

\begin{table*}[ht]
  \centering
  \caption{NDCG@10 on TREC-DL 2019/2020 and four BEIR subsets. \underline{\textbf{Bold-underlined}}: best; \underline{underlined}: second-best.}
  \label{tab:tab3}
  \resizebox{\textwidth}{!}{ 
  \begin{tabular}{l|l|cccccc|c}\toprule
    LLM & methods & COVID  & NFCorpus & DBPedia & SciFact & DL-19 & DL-20 & Avg \\
    \midrule
    - & BM25 & 0.595  & 0.322 & 0.318 & 0.679  & 0.506 & 0.480 & 0.483\\
    \midrule
    \multirow{6}{*}{Flan-T5-XL} & Pointwise-RG & 0.698  & 0.331 & 0.273 & 0.553  & 0.650 & 0.636 & 0.524  \\
    & Pairwise-Bubblesort & 0.763 & \underline{\textbf{0.359}} & \underline{0.432} & \underline{\textbf{0.734}} & 0.683 & 0.662 & \underline{0.606}  \\
    & Setwise-Heapsort & 0.757  & 0.352 & 0.428 & 0.677  & 0.692 & \underline{0.678} & 0.597  \\
    & Listwise & 0.650  & 0.334 & 0.366 & 0.694 & 0.568 & 0.547 & 0.527  \\
    & \textbf{RefRank-Single(1)} & \underline{0.777}  & 0.354 & 0.423 & 0.666 	& \underline{0.697} & 0.659 & 0.596  \\
    & \textbf{RefRank-Multiple(4)} & \underline{\textbf{0.785}}  & \underline{0.357} & \underline{\textbf{0.443}} & \underline{0.695}  & \underline{\textbf{0.702}} & \underline{\textbf{0.680}} & \underline{\textbf{0.610}}  \\
    \midrule
    \multirow{6}{*}{Flan-T5-XXl} & Pointwise-RG & 0.691  & 0.322 & 0.305 & 0.623  & 0.642 & 0.632 & 0.536  \\
    & Pairwise-Bubblesort & 0.733  & \underline{\textbf{0.363}} & \underline{0.421} & \underline{\textbf{0.756}} & 0.679 & 0.681 & 0.605  \\
     & Setwise-Heapsort & 0.752  & 0.346 & 0.402 & 0.726  & \underline{0.706} & \underline{0.688} & 0.603\\
    & Listwise & 0.664  & 0.344 & \underline{\textbf{0.441}} & 0.736  & 0.660 & 0.637 & 0.580  \\
    & \textbf{RefRank-Single(1)} & \underline{0.769}  & 0.352 & 0.406 & 0.731 & \underline{\textbf{0.707}} & 0.682 & \underline{0.608} \\
    & \textbf{RefRank-Multiple(4)} & \underline{\textbf{0.783}}  & \underline{0.358} & 0.415 & \underline{0.746}  & 0.702 & \underline{\textbf{0.699}} & \underline{\textbf{0.617}} \\
    \midrule
    \multirow{6}{*}{Llama-3.1-8B} & Pointwise-RG & 0.734  & 0.335 & 0.307 & 0.561  & 0.617 & 0.580 & 0.522  \\
    & Pairwise-Bubblesort & 0.768  & \underline{\textbf{0.351}} & \underline{\textbf{0.396}} & \underline{\textbf{0.747}} & 0.652 & \underline{0.606} & \underline{0.586}  \\
    & Setwise-Heapsort & 0.794  & 0.336 & 0.371 & \underline{0.731}  & 0.662 & 0.600 & 0.582\\
    & Listwise & \underline{\textbf{0.812}}  & 0.344 & 0.384 & 0.691  & 0.681 & \underline{\textbf{0.678}} & \underline{\textbf{0.598}}  \\
    & \textbf{RefRank-Single(1)} & 0.790  & 0.343 & 0.381 & 0.707 & \underline{0.683} & 0.597 & 0.583 \\
    & \textbf{RefRank-Multiple(4)} & \underline{0.809}  & \underline{0.348} & \underline{0.392} & 0.727  & \underline{\textbf{0.704}} & \underline{0.606} & \underline{\textbf{0.598}} \\
    \bottomrule
  \end{tabular}
  }
\end{table*}

\section{Ablation Studies}

We conduct two controlled ablations to quantify (i) the sensitivity of RefRank to the choice of a single reference document and (ii) the marginal utility of aggregating an increasing number of reference documents.  
All ablations share the identical pipeline and hyper-parameters used in the main experiments (\S~\ref{sec4}), ensuring that any observed differences are solely attributable to the factor under investigation.

\subsection{Reference choice stability}
\label{sec:ablation_single}

To assess the sensitivity of RefRank to the selection of a single reference document, we conduct an ablation study using the top-5 passages retrieved by BM25 for each query. Specifically, we sequentially select the k-th ranked passage (k=1,…,5) as the sole reference signal, resulting in five variants: \textbf{Single(1)} through \textbf{Single(5)}. No other components of the ranking pipeline are modified. We evaluate the average NDCG@10 across six datasets: TREC-DL 2019, TREC-DL 2020, and four subsets from BEIR.

\begin{figure}[ht]
  \centering
  \begin{subfigure}[b]{0.49\linewidth}
    \includegraphics[width=\linewidth]{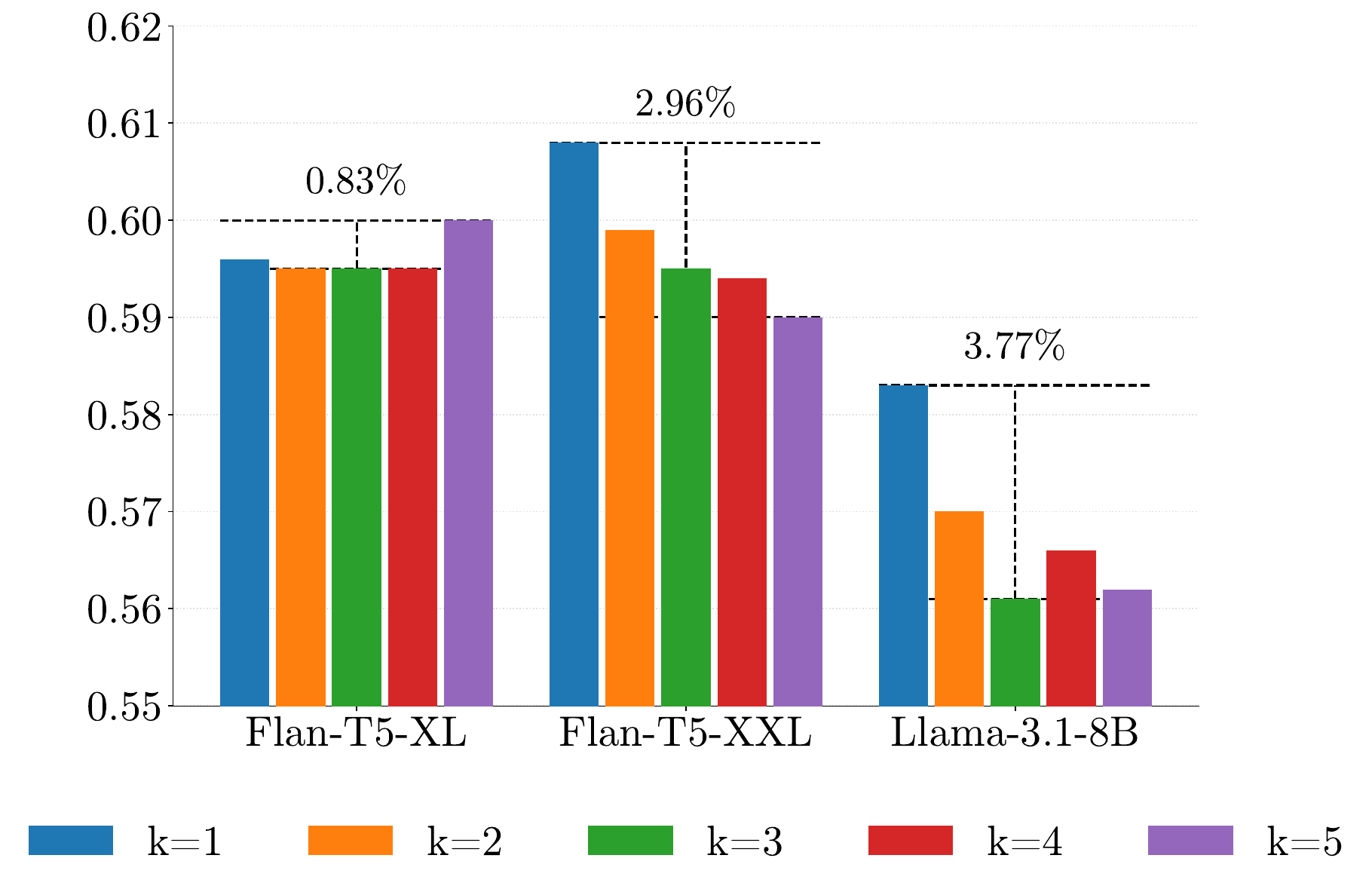}
    \caption{Reference document selection sensitivity}
  \end{subfigure}
  \hfill
  \begin{subfigure}[b]{0.49\linewidth}
    \includegraphics[width=\linewidth]{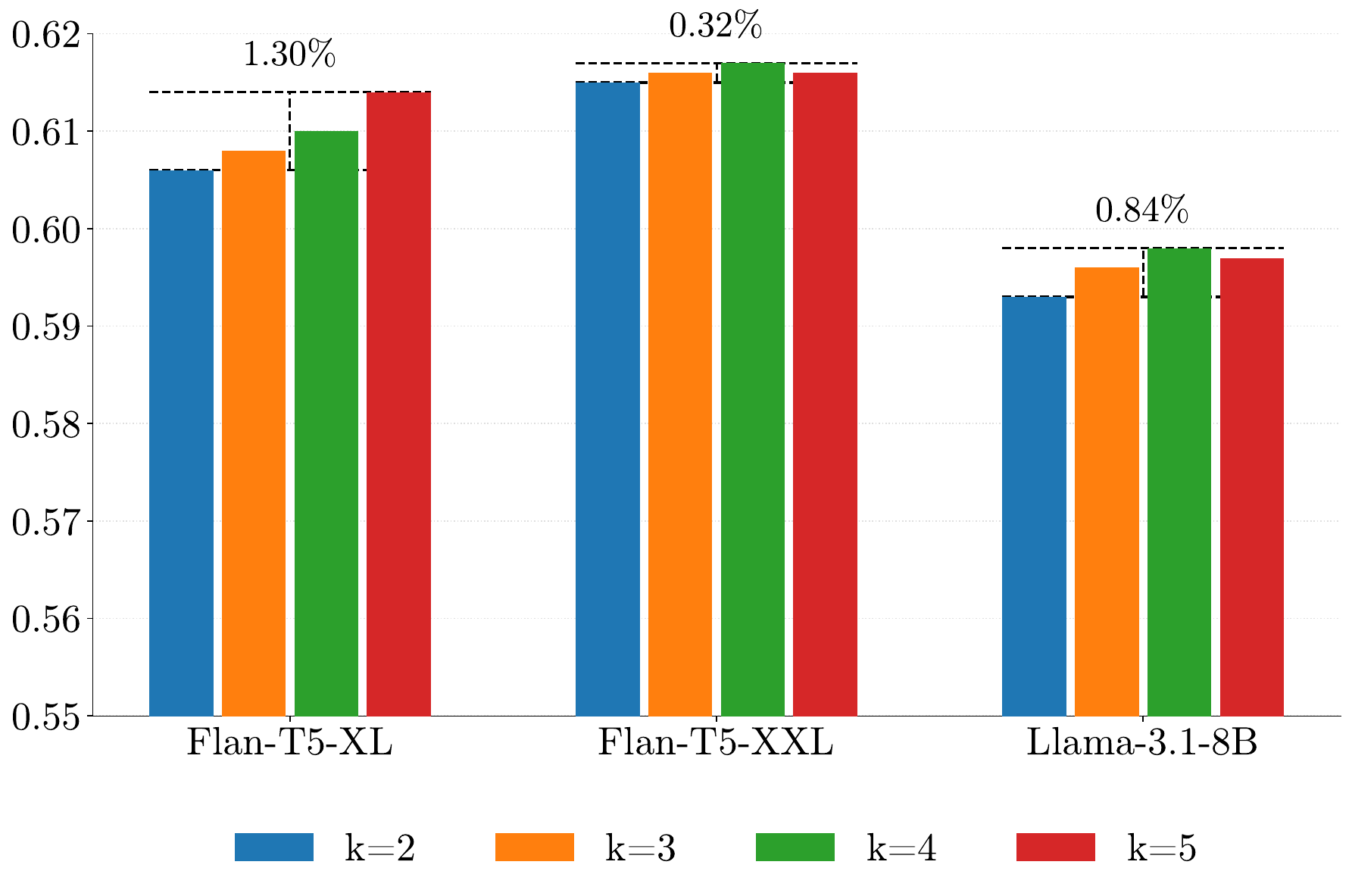}
    \caption{Utility of aggregating multiple reference documents}
  \end{subfigure}

  \caption{Ablation study results on (a) reference document selection sensitivity and (b) utility of aggregating multiple reference documents. Performance is measured by average NDCG@10 across six datasets. The left plot shows minimal variation when using a single reference document ranked at position $k$ (Single($k$)), indicating robustness to reference choice. The right plot demonstrates consistent improvements when aggregating multiple references (Multiple($k$)).}
  \label{fig:absolution}
\end{figure}

Figure~\ref{fig:absolution} quantifies how sensitive RefRank-Single(1) is to the position of the reference document that is injected.
Across the three model families the standard deviation of NDCG@10 never exceeds 0.008 and the worst-case swing is only 3.77\% (Llama-3.1-8B: 0.561--0.583).
Such a tight band implies that any passage in the top-5 already contains enough lexical and semantic cues to anchor the pairwise comparisons; the downstream transformer is therefore able to compensate for small topical drifts without reranking failures.
This observation is practically important: it frees engineers from exhaustive tuning of the reference selector and guarantees stable deployment when the exact rank of the presumed-best passage is uncertain.

Although the variance is small, the relationship between the rank position of the chosen reference passage and effectiveness is nearly monotonic: selecting the top-1 BM25 passage as reference yields the highest average NDCG@10 on Flan-T5-XXL and Llama-3.1-8B.
The gain is statistically marginal when averaged over the six datasets, yet the direction is consistent: the top-1 BM25 passage always yields the highest—or tied-highest—mean NDCG@10. This indicates that the initial ranker already surfaces a slightly superior “pivot”. Consequently, we adopt top-1 as the default reference: it incurs zero extra cost, delivers the best expected performance, and keeps the implementation trivially simple.


\subsection{Utility of score aggregation}
\label{sec:ablation_multiple}
 
To investigate the marginal utility of aggregating an increasing number of reference documents, we systematically expand the reference set size from $k=2$ to $k=5$, and compute the average likelihood probability assigned by the LLM to each candidate passage. The resulting variants are denoted as \textbf{Multiple(2)} to \textbf{Multiple(5)}.

As illustrated in Figure~\ref{fig:absolution}, the performance variation across different values of \(k\) is minimal across all models. Specifically, for Flan-T5-XL, the maximum and minimum scores are 0.614 and 0.606, respectively, yielding a difference of 0.008 (a 1.30\% relative change). For Flan-T5-XXL, the scores range from 0.615 to 0.617, with a difference of only 0.002 (a 0.32\% relative change). Similarly, Llama-3.1-8B exhibits a variation of 0.005 (0.84\%) between its maximum (0.598) and minimum (0.593) scores.

Importantly, all aggregated variants consistently outperform the single-reference baseline. Moreover, we observe that using the top-4 references (\(k=4\)) yields a local maximum for both Llama-3.1-8B and Flan-T5-XXL. Therefore, when computational resources permit, selecting the top-4 references for aggregation represents a reasonable and effective choice in practice.

\section{Conclusion}
\label{sec:concl}

This work reframes LLM reranking calibration as an anchoring rather than a post-hoc regression problem.
By reusing the top-1 retrieval passage as a fixed, query-specific anchor, RefRank converts any generative pointwise scorer into a contrastive comparator without extra parameters, training, or prompt engineering.
Across six standard test collections, the method establishes a new zero-shot NDCG@10 peak for Flan-T5 and Llama-3.1-8B at the cost of negligibly additional latency, and the gain is additive to existing prompt tricks.

Our immediate agenda is to learn a query-difficulty-aware meta-controller that dynamically expands or shrinks the anchor set, and to extend RefRank to multi-modal and streaming retrieval where anchors can be images, audio, or on-the-fly generated pseudo-documents.
By decoupling high-quality calibration from model size and inference budget, RefRank democratizes accurate reranking for resource-constrained edge and real-time search scenarios.

\bibliography{RefRank}
\bibliographystyle{tmlr}


\end{document}